\documentclass[letterpaper,prl,twocolumn,showpacs,showkeys,preprintnumbers,amsmath,amssymb,superscriptaddress]{revtex4}
\usepackage{graphicx}
\usepackage[figuresright]{rotating}
\usepackage{amsmath}

\begin{document}
\title{The $^{\bf 21}\mbox{Na}(\mbox{p},\boldsymbol{\gamma})^{\bf 22}\mbox{Mg}$
Reaction and Oxygen-Neon Novae}
\author{S. Bishop}
\affiliation{Simon Fraser University, Burnaby, B.C., Canada}
\author{R. E. Azuma}
\affiliation{University of Toronto, Toronto, Ont., Canada}
\author{L. Buchmann}\affiliation{TRIUMF, Vancouver, B.C., Canada}
\author{A. A. Chen}\altaffiliation{Present address: McMaster Univ., Hamilton, Ont., Canada}
\affiliation{Simon Fraser University, Burnaby, B.C., Canada}
\author{M. L. Chatterjee}
\affiliation{Saha Institute of Nuclear Physics, Calcutta, India}
\author{J. M. D'Auria}\email{dauria@sfu.ca}
\affiliation{Simon Fraser University, Burnaby, B.C., Canada}
\author{S. Engel}
\affiliation{Ruhr-Universit\"{a}t, Bochum, Germany}
\author{D. Gigliotti}
\affiliation{University of Northern British Columbia, Prince George, B.C., Canada}
\author{U. Greife}
\affiliation{Colorado School of Mines, Golden, CO, USA}
\author{M. Hernanz}\affiliation{Institut d'Estudis Espacials de Catalunya, CSIC/UPC,
Barcelona, Spain}
\author{D. Hunter}\affiliation{Simon Fraser University, Burnaby, B.C., Canada}
\author{A. Hussein}
\affiliation{University of Northern British Columbia, Prince George, B.C., Canada}
\author{D. Hutcheon}\affiliation{TRIUMF, Vancouver, B.C., Canada}
\author{C. Jewett}\affiliation{Colorado School of Mines, Golden, CO, USA}
\author{J. Jos{\'{e}}}\affiliation{Institut d'Estudis Espacials de Catalunya, CSIC/UPC,
Barcelona, Spain} \affiliation{Universitat Polit\'ecnica de
Catalunya, Barcelona, Spain}
\author{J. King}\affiliation{University of Toronto, Toronto, Ont., Canada}
\author{S. Kubono}\affiliation{University of Tokyo, Tokyo, Japan}
\author{A. M. Laird}\affiliation{TRIUMF, Vancouver, B.C., Canada}
\author{M. Lamey}\affiliation{Simon Fraser University, Burnaby, B.C., Canada}
\author{R. Lewis}\affiliation{Yale University, New Haven, CT, USA}
\author{W. Liu}\affiliation{Simon Fraser University, Burnaby, B.C., Canada}
\author{S. Michimasa}\affiliation{University of Tokyo, Tokyo, Japan}
\author{A. Olin}\affiliation{TRIUMF, Vancouver, B.C., Canada}
\affiliation{University of Victoria, Victoria, B.C., Canada}
\author{D. Ottewell}\affiliation{TRIUMF, Vancouver, B.C., Canada}
\author{P. D. Parker}\affiliation{Yale University, New Haven, CT, USA}
\author{J. G. Rogers}\affiliation{TRIUMF, Vancouver, B.C., Canada}
\author{F. Strieder}\affiliation{Ruhr-Universit\"{a}t, Bochum, Germany}
\author{C. Wrede}\affiliation{Simon Fraser University, Burnaby, B.C., Canada}
%
%
\pacs{25.60.-t, 26.30.+k}
\keywords{nucleosynthesis; resonance reaction; radioactive beam;
radiative proton capture; thick target yield; white dwarf; novae}
\date{\today}
\begin{abstract}
The $^{21}\mbox{Na}(\mbox{p},\gamma)^{22}\mbox{Mg}$ reaction is
expected to play an important role in the nucleosynthesis of
$^{22}\mbox{Na}$ in Oxygen-Neon novae. The decay of
$^{22}\mbox{Na}$ leads to the emission of a characteristic 1.275
MeV gamma-ray line.  This report provides the first direct
measurement of the rate of this reaction using a radioactive
$^{21}\mbox{Na}$ beam, and discusses its astrophysical
implications. The energy of the important state was measured to be
E$_{c.m.}$= 205.7 $\pm$ 0.5 keV with a resonance strength
$\omega\gamma = 1.03\pm0.16_{stat}\pm0.14_{sys}$~meV.

\end{abstract}
\maketitle
%

 The synthesis of light and intermediate-mass elements
 can take place through radiative proton captures on
 unstable nuclei during explosive stellar events.
 One astrophysical site where such processes can
 occur involves classical novae, stellar explosions
 powered by thermonuclear runaways on accreting
 ONe or CO white dwarf stars~\cite{wiescher,forty,jose1}.
 In this hydrogen burning
 process, nuclear activity involves different cycles,
 depending on the nova type and on the temperatures
 achieved during the explosion. A predominant nuclear
 activity in ONe novae takes place in the NeNa cycle,
 initiated by radiative proton captures on the abundant
 seed nuclei $^{20}$Ne.

 Nucleosynthesis in the NeNa cycle during nova outbursts
 leads to the synthesis of the astronomically important, but
 unstable $^{22}$Na nucleus. Its $\beta$-decay ($t_{1/2}$ = 2.6 yr)
 leads to the emission of a 1.275 MeV $\gamma$-ray,
 following population of the first excited state of
 $^{22}$Ne. In fact, this $\gamma$-ray is an ideal
 observable for nova events as first suggested by
 Clayton \& Hoyle ~\cite{clayton}. Thus far, observational
 searches performed with NASA's COMPTEL on-board CGRO
 satellite of five ONe novae have not found this
 $\gamma$-ray signature~\cite{iyudin}. Whereas the inferred
 upper limits are in agreement with recent results
 from ONe nova models~\cite{forty,jose1}, the reduction of the
 nuclear uncertainties associated with the main
 reactions involved in the synthesis of $^{22}$Na
 is critically important in order to predict how
 much $^{22}$Na can be produced in a typical nova
 event, and at what distance a nova explosion is expected
 to provide a detectable flux of $\gamma$-rays.

 Another aspect that stresses the astronomical
 interest of $^{22}$Na relies on the identification
 of presolar grains likely condensed in the ejecta
 from nova outbursts.
 Traditionally, they have been identified by low $^{20}$Ne/$^{22}$Ne ratios
 (where $^{22}$Ne is attributed to in-situ $^{22}$Na decay).
 A $^{22}$Na/C ratio of $9 \times 10^{-6}$ ~\cite{nichol}
 has been determined recently in the graphite
 grain KFB1a-161, in which other isotopic ratios resemble those
  found in the envelopes ejected by nova outbursts. Again, a more
  accurate knowledge of reactions in the synthesis
  of $^{22}$Na in novae would further assist in identifying
  presolar grains from novae and for tuning models accordingly.

\begin{figure}[t!]

\scalebox{.23}{\includegraphics{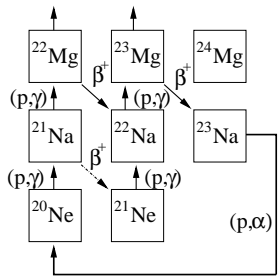}}

\begin{center}
\caption{\label{fig:cycles}The  combined``cold''and ``hot'' NeNa
reaction cycles. The isotope $^{21}\mbox{Na}$ will either
$\beta$-decay into $^{21}\mbox{Ne}($the ``cold'' NeNa cycle) or
capture a proton leading to $^{22}\mbox{Mg}$ (the ``hot'' cycle)
depending upon the temperature and the reaction rate.}
\end{center}
\end{figure}

Synthesis of $^{22}$Na in novae takes place following two possible
reaction paths (Fig.~\ref{fig:cycles}): in the first (``cold''
NeNa cycle), $^{21}$Na forms from the seed $^{20}$Ne which then
leads to
  ${}^{21}$Na($\beta^+$)${}^{21}$Ne(p,$\gamma$)${}^{22}$Na;
in the second path, associated with higher temperatures (``hot''
NeNa cycle), proton-capture on $^{21}$Na dominates over its
$\beta$-decay, followed by
   ${}^{21}$Na(p,$\gamma$)${}^{22}$Mg($\beta^+$)${}^{22}$Na.
   There is little net mass flow
from $^{22}\mbox{Mg}$ to $^{23}\mbox{Al}$ due to the low Q-value
for photodisintegration of $^{23}\mbox{Al}$~\cite{wiescher2}.
 Current models of ONe novae indicate that the unknown rate of
 $^{21}$Na(p,$\gamma$) is the main source of uncertainty associated
 with calculating the amount of $^{22}$Na in nova outbursts~\cite{jose2,coc}.
 The purpose of this paper is to report on the first direct measurement
 of this rate.

Under nova conditions the capture reaction rate,
$N_\text{A}\langle\sigma v\rangle$, is expected to be dominated by
one or more narrow resonances. Each resonance contributes to the
reaction rate in direct proportion to its resonance strength,
$\omega\gamma$, and depends exponentially on the resonance energy,
 $E_\text{R}$.  In units of $\mbox{{cm}}^3\mbox {s}^{-1}\mbox{ mol}^{-1}$,
 it is given by,
\begin{equation}
N_\text{A}\langle\sigma v \rangle=1.54\times 10^{11} (\mu
T_{9})^{-3/2} \omega\gamma
\,\exp\left[{-11.605\frac{E_\text{R}}{T_{9}}}\right]\;,
\label{eq:rate}
\end{equation}
with $N_\text{A}$ Avogadro's number, $\mu$ the reduced mass in u,
$T_{9}$ the temperature in units of GK, $\langle\sigma v \rangle$
the thermally averaged nuclear cross section, and $\omega\gamma$
and $E_\text{R}$ in MeV~\cite{fowler}. The narrow resonance thick
target yield, $Y$, at maximum is~\cite{rolfsbook},
\begin{equation}
Y =
\frac{\lambda^2}{2}\frac{M+m}{m}\omega\gamma\left(\frac{dE}{dx}\right)^{-1}\;,
\label{eq:yield}
\end{equation}
with $\lambda$ the centre-of-mass de Broglie wave length, $M$ the
(heavy) projectile nucleus mass, $m$ the (light) target nucleus
mass, and $\frac{dE}{dx}$ the energy loss per atom/cm$^{2}$(lab).
Thus, measurement of the maximum thick target yield can determine
the resonance strength, $\omega\gamma$.

\begin{figure}[h]
\scalebox{.32}{\includegraphics{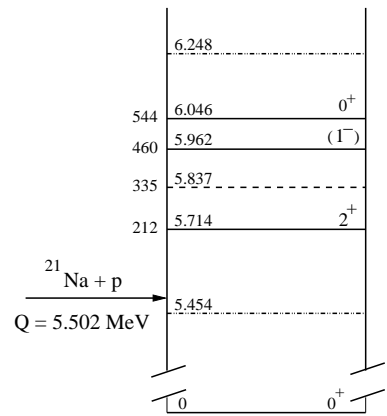}}
\caption{\label{fig:levels}The $^{22}\mbox{Mg}$ level scheme of
those states of astrophysical interest for ONe nova, shown with
solid lines~\cite{bateman,kubono}.  The numbers on the far left
denote centre-of-mass energies (E$_{x}$- Q) in units of keV. The
state at 5.837 MeV was
  observed once but not confirmed in other studies~\cite{bateman,kubono,chen}. }
\end{figure}

Figure~\ref{fig:levels} shows the $^{22}\mbox{Mg}$ level
scheme~\cite{bateman,kubono}. Calculation of the Gamow window
indicates that the 212 keV, $\ell=0$ resonance will be the
dominant contributor (as compared to other higher resonances and
direct capture) to the
$^{21}\mbox{Na}(\mbox{p},\gamma)^{22}\mbox{Mg}$ reaction at all
nova temperatures from 0.2 to 0.35~GK. We report here a
measurement of the strength, $\omega\gamma$, and a revised energy,
$E_\text{R}$, for this resonance in the
$^{21}\mbox{Na}(\mbox{p},\gamma)^{22}\mbox{Mg}$ reaction.

The experiment was carried out at the TRIUMF-ISAC radioactive
beams facility located in Vancouver, Canada. Fifteen $\mu$A of 500
MeV protons bombarded a thick target of SiC resulting in an
intense ($\sim10^{9}~\mbox{s}^{-1}$), pure ($\sim100$\%) $^{21}$Na
beam extracted from a surface ion source and a high resolution
mass analyzer~\cite{domb}. It was accelerated using the new ISAC
linear accelerator, resulting in beams with energies variable from
0.15 to 1.5 MeV/u~\cite{lax}. The study was performed using
inverse kinematics with the DRAGON (Detector of Recoils And Gammas
Of Nuclear reactions) facility. DRAGON consists of a windowless
gas target (effective length of 12.3 cm) surrounded by a gamma
array (30 units of BGO), and followed by a two-stage, recoil mass
separator, 21 m in length (from target centre to focal plane).
Separation of the rare recoil from more intense beam is achieved
using magnetic and electric dipoles. Following an initial
selection of a single (optimal) charge state~\cite{liu} in the
first magnetic dipole, energy dispersion in the electric dipole
allows mass separation, and the process is repeated in the second
stage. A DSSSD (Double Sided Silicon Strip Detector) was used at
the focal plane of DRAGON to detect the $^{22}\mbox{Mg}$ recoils.
A more complete description of DRAGON can be found
elsewhere~\cite{jda,hutch}.

\begin{figure}
\scalebox{.36}{\includegraphics{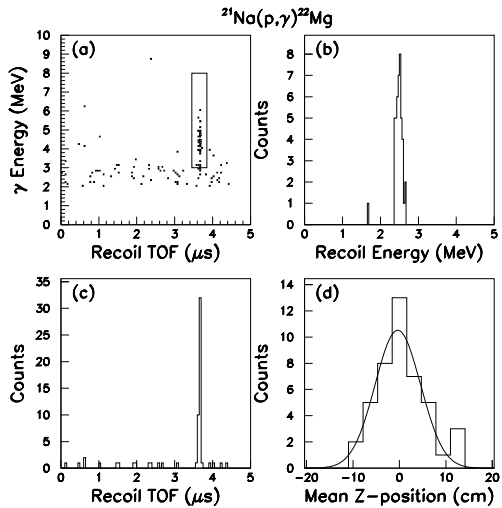}}
\caption{\label{fig:tof} Resonant-capture spectra for a $^{21}$Na
beam energy of 220 keV/u. (a) Valid events enclosed by a
two-dimensional-cut box above a background of random-coincidence
events,(b) the recoil-energy distribution of the events in the
DSSSD, selected by the box in (a), (c) the  recoil TOF
distribution for events above the $\gamma$-ray threshold energy,
(d) distribution of box-selected $\gamma$-ray events observed in
the BGO array along the target length, with a Gaussian fit.}
\end{figure}

A radioactive beam of $^{21}\mbox{Na}$ (q=5$^{+}$) at typical
intensities up to $5\times 10^{8}~\mbox{s}^{-1}$ was delivered to
the DRAGON hydrogen gas target (4.6 Torr). The gas target received
a total of $\sim 10^{13}$ $^{21}\mbox{Na}$ atoms for this study.
Data taking was done in both singles and coincidence modes; the
coincidence mode required a ``start'' timing signal from the
$\gamma$-array in coincidence with a ``stop'' timing signal from
the DSSSD. Figure~\ref{fig:tof} shows resonant-capture spectra for
a beam energy of 220 keV/u. Counts within the box in
Fig.~\ref{fig:tof}a were considered to be valid capture events.
Their recoil energy distribution is presented in
Fig.~\ref{fig:tof}b. Fig.~\ref{fig:tof}c is the recoil
time-of-flight spectrum for events satisfying the cut on gamma-ray
energy. The distribution of the hit BGO detector position along
the beam axis (Fig.~\ref{fig:tof}d) shows that the resonance was
near the centre of the gas target at beam energy 220 keV/u
($E_{c.m.}$= 211 keV).

\begin{figure}[h]
\begin{minipage}{66mm}
\scalebox{.34}{\includegraphics{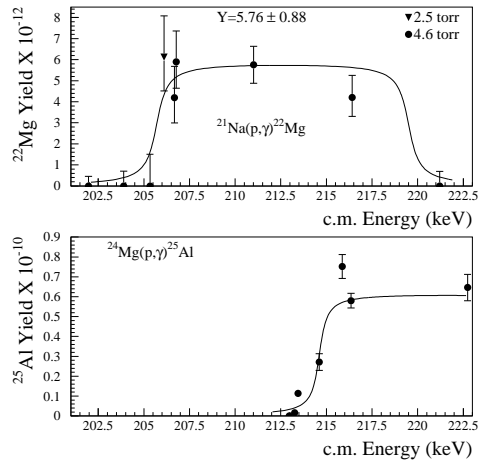}}
\end{minipage}
\caption{\label{fig:yield}The upper panel displays the thick
target yield data for the
$^{21}\mbox{Na}(\mbox{p},\gamma)^{22}\mbox{Mg}$ reaction, with the
solid line showing the nominal target thickness for 4.6 Torr.
Yield of the $^{24}$Mg(p,$\gamma$)$^{25}$Al reaction for the
resonance at E$_{c.m.}$=214 keV, used for beam energy calibration,
is displayed in the lower panel. Statistical errors only are
displayed in both.}
\end{figure}

The beam energies were measured by adjusting the field of the
first magnetic dipole in the separator so as to position the beam
on the ion-optical axis at an energy-dispersed focus. Using the
design bending radius of the dipole (1 m), it was possible to
calculate beam energy in terms of the dipole field. The expected
relationship was confirmed by measuring a number of known
resonances with stable beams.  The lower panel of
Fig.~\ref{fig:yield} shows the yield curve for one of these
studies, the $^{24}\mbox{Mg}(\mbox{p},\gamma)^{25}\mbox{Al}$
reaction, demonstrating our agreement (inflection point of 214.4
$\pm$ 0.5 keV) with the literature resonance energy of 214.0 $\pm$
0.1 keV ~\cite{uhrm}. As shown in the upper panel, we find the
energy (inflection point) for the $^{21}$Na(p,$\gamma$)$^{22}$Mg
resonance to be 205.7 $\pm$ 0.5, and not 212 keV (see
Fig.~\ref{fig:levels}), the difference between the Q value and the
level excitation energy, 5713.9 $\pm$ 1.2 keV~\cite{endt}. Given
that the latter value is based upon a direct gamma de-excitation
measurement of the 5713.9 keV level, this disagreement could be
explained by a modified mass excess for $^{22}$Mg; our data imply
a value of -403.2 $\pm$ 1.3 keV rather than
-396.8~keV~\cite{hardy}.

Figure~\ref{fig:yield} (upper panel) shows the thick target yield
curve corrected/scaled for various factors listed in
Table~\ref{tab:error}. The efficiency of the BGO array as a
function of $\gamma$-ray energy and resonance position in the
target was calculated using the GEANT program ~\cite{gigl,gig2}.
The variation of resonance position with beam energy resulted in
the following calculated efficiencies: 45\% for 202~keV $\leq$ E
$\leq$ 207~keV, 48\% at 211~keV, and 46\% above 216~keV. The
systematic error was deduced from values of the array efficiency
measured with stable beam reactions. The separator transmission
(98\%)~\cite{hutch} and DSSSD detection efficiency
(99\%)~\cite{wrede} were determined separately, and the fraction
of the charge state selected (44\%) was measured with a $^{24}$Mg
beam of 220 keV/u. At 4.6 Torr, charge state equilibrium in
H$_{2}$ gas was measured to be attained within 2 mm ~\cite{liu}.
The energy loss in the target (4.6 Torr) was measured to be~14.4
keV/u (lab) or $8.18\times10^{-14}\text {eV/(atom/}\text{cm}^2$),
in agreement with SRIM~\cite{srim}.

The data of Fig.~\ref{fig:yield} (upper panel) were obtained by
maximum likelihood combination of several runs at each
energy~\cite{bishop}. The error bars on the zero counts seen at
off-resonance energies are 68$\%$ confidence limits.
Table~\ref{tab:error} presents a summary of systematic errors.
Using Eq.~\ref{eq:yield} and only the mid-target data point (211
keV), a yield of $(5.76\pm 0.88)\times 10^{-12}$ per incident
$^{21}$Na, results in a resonance strength of $\omega\gamma
=1.03\pm0.16_{stat}\pm0.14_{sys}$~meV.

\begin{table}
\caption{Summary of  systematic errors.}
\begin{ruledtabular}
\label{tab:error}
\begin{tabular}{ccccc}
Factors & Value  &Syst.Error($\%$)  \\
\hline
BGO array efficiency (@211 keV) & 0.48   &   12  \\
Separator transmission     &   0.98    &  2            \\
DSSSD efficiency        &    0.99    &  1      \\
Charge state fraction    &      0.44    & 3     \\
Integrated beam (@211 keV)  &  $3.62\times10^{13}$   &  4          \\
dE/dx ($\text{eV/(atom/}\text{cm}^2))_{lab}$  &
$8.18\times10^{-14}$ &     5
\\
\end{tabular}
\end{ruledtabular}

\end{table}

\begin{figure}
\scalebox{.33}{\includegraphics{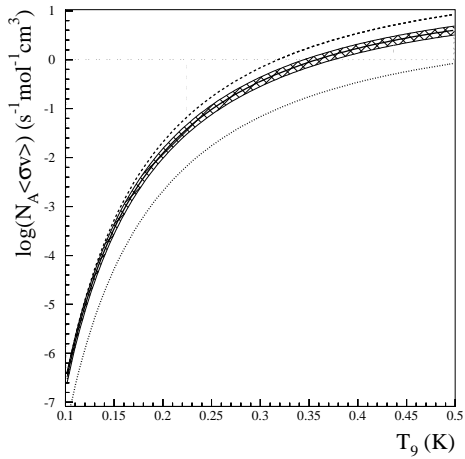}}
\caption{\label{fig:rate}The stellar rate for the
$^{21}\mbox{Na}(\mbox{p},\gamma)^{22}\mbox{Mg}$ reaction using Eq.
2 with typical novae temperatures and our measured values for
$\omega\gamma$ and $E_\text{R}$= 0.206 MeV (solid line with
hatched area reflecting errors), in comparison with other works;
upper curve~\cite{bateman} and lower curve~\cite{jose2}.}
\end{figure}

The effect of these results on the calculated stellar reaction
rate is shown in Fig.~\ref{fig:rate}. The  rate is reduced over
that determined by shell model calculations of $\omega\gamma$ as
reported in~\cite{bateman}, and enhanced over that found
in~\cite{jose2}. An analysis of the impact of the new measurements
on the synthesis of $^{22}$Na in novae was performed. A new model
of a nova outburst, using an ONe white dwarf of 1.25 solar mass,
has been computed from the onset of accretion up to the explosion
and ejection stages, by means of a spherically symmetric,
implicit, hydrodynamic code, in Lagrangian formulation
(see~\cite{jose1} for details). Results have been compared with a
model evolved with the previous prescription of the
$^{21}$Na(p,$\gamma$)$^{22}$Mg rate~\cite{jose2}. As a result of
the higher contribution of the 5.714 MeV level
(Fig.~\ref{fig:rate}), a slightly lower amount of $^{22}$Na (a
mean mass fraction of $2.8 \times 10^{-4}$, compared with the
previous estimate of $3.5 \times 10^{-4}$) is found. The small
decrease in the $^{22}$Na yield results from the fact that
increasing the proton capture rate on $^{21}$Na favours the
synthesis path through
${}^{21}$Na(p,$\gamma$)${}^{22}$Mg($\beta^+$)${}^{22}$Na, hence
reducing the role of the alternative
${}^{21}$Na($\beta^+$)${}^{21}$Ne(p,$\gamma$)${}^{22}$Na path. In
these newly derived conditions of increased proton capture on
$^{21}$Na, $^{22}$Na production takes place earlier in the
outburst, at a time when the envelope has not yet significantly
expanded and cooled down (contrary to the case when a lower
$^{21}$Na(p,$\gamma$) rate is adopted), and hence the temperature
in the envelope is still high enough to allow proton captures on
$^{22}$Na, that reduce its final content in the ejecta.

 Up to now, $\gamma$-ray flux
determinations were limited by a large uncertainty in the
   the $^{21}$Na(p,$\gamma$) and $^{22}$Na(p,$\gamma$)
   rates, which translated into an overall
   uncertainty in the $^{22}$Na yields of a factor of $\sim 3$.
   The maximum detectability distance was, accordingly, uncertain by
   a factor of $\sim 2$. Such uncertainty, mainly due to the previously
 unknown reaction rate,
    has been largely reduced with the present
experimental determination of $\omega \gamma= 1.03 \pm 0.16_{stat}
\pm 0.14_{sys}$ meV. These results provide a firmer basis for
predictions of the
   expected gamma-ray signature at 1.275 MeV associated with $^{22}$Na decay
   in ONe novae, and confirm the previous determination of 1 kiloparsec
   for a typical ONe nova ~\cite{gomez,hernanz} observed with ESA's
   (European Space Agency) INTEGRAL spectrometer, SPI.
   Furthermore, the smaller uncertainty in the rate also indicates that
   the predicted $^{22}$Na yields are not in conflict with
   the upper limits derived from several observational searches.

Financial support from the Natural Sciences and Engineering
Research Council of Canada, the Deutsche Forschungsgemeinschaft
(DFG GR 1577/3), the Department of Energy grant
(DE-FG03-93ER40789) of the U.S.A., and from TRIUMF is gratefully
acknowledged.

\bibliography{jose}

\end{document}